\documentstyle[12pt]{article}
\newcommand{\f}{\begin{equation}}
\newcommand{\ff}{\end{equation}}
\newcommand{\blankmm}{\vskip .2cm}
\newcommand{\blankline}{\vskip .3cm}
\newcommand{\blankblock}{\vskip 3cm}
\newcommand{\fff}{{\bf\Psi}}
\newcommand{\half}{{1\over2}}

\newcommand{\pl}{Phys. Lett.\ }
\newcommand{\cmp}{Commun. Math. Phys.\ }
\newcommand{\np}{Nucl. Phys.\ }
\newcommand{\Tr}{{\rm Tr}}

\begin{document}
\begin{titlepage}

\begin{flushright}
UM-P-93/107 \\
hep-th/9403061 \\
February 1994 \\
\end{flushright}

\blankblock
\centerline{\LARGE Covariant Phase Space Formulations of Superparticles}
\blankmm
\centerline{\LARGE and Supersymmetric WZW Models}
\blankline
\blankline
\blankline
\centerline{G. Au\footnote{gau@tauon.ph.unimelb.edu.au}
     and B. Spence\footnote{spence@tauon.ph.unimelb.edu.au}}
\centerline{School of Physics}
\centerline{University of Melbourne}
\centerline{Parkville 3052}
\centerline{Australia}
\blankline

\begin{abstract}
We present new covariant phase space formulations of superparticles
moving on a group manifold, deriving the fundamental Poisson
brackets and current algebras. We show how these formulations
naturally generalise to the supersymmetric Wess-Zumino-Witten models.
\end{abstract}

\vfill
\end{titlepage}
\newpage
\section{Introduction}

The Wess-Zumino-Witten (WZW) models \cite{witten}
are fundamental rational conformal field
theories, and have a rich structure which has occasioned much interest.
As the WZW models
are expected to possess most of the characteristic features of rational
conformal field theories, they are of considerable
interest in the study of conformal field theory and string theory in general.
In particular, there has been some inquiry of late into the phenomenon of
the emergence of quantum groups in WZW models (\cite{blok} - \cite{hasie}).
WZW models are also of interest due to the fact
that Hamiltonian reduction of these models gives another
important class of integrable two-dimensional systems - the Toda theories
(see \cite{irish} for a review).

A new covariant phase space description of the non-supersymmetric
WZW model has been found recently \cite{papasptwo}. This description makes
possible
a particularily simple derivation
of the fundamental Poisson brackets of the theory. It is also clear from
this formulation that the differences between it and
other approaches in this field
arise due
to the differing topologies of the phase spaces involved. An appealing
feature of the formulation in reference \cite{papasptwo}
is that the parameterisation of the solutions of the theory given there is such
as
to guarantee that the covariant phase space is
diffeomorphic to the Hamiltonian phase space, and thus this covariant phase
space formulation gives agreement with the usual Hamiltonian methods.
With regard to the further development of the formulation of this approach,
as well as to the various
applications of supersymmetric WZW models in superstring theories,
it is of interest to consider the question of whether
one can generalise this covariant phase space formulation to the
{\it supersymmetric}
WZW models. In this paper, we will present such formulations.

In section 2 we will discuss superparticles moving upon group manifolds.
These
systems share many of the important features of the supersymmetric WZW models.
The WZW models are then discussed in section 3, taking advantage of
the results of the earlier particle discussion. We
show how the full current algebras for all these models arise naturally
in our approach. We then show that the topological issues which arose in the
bosonic case also are found here, with the same resolutions.
We finish with some concluding
remarks in section 4.

\section {Superparticles on Group Manifolds}
An analysis of the one-dimensional supersymmetric non-linear sigma models has
been given in ref. \cite{coles}. The class of such models defined on group
manifolds is more general than those considered here - as we are interested
in the two-dimensional supersymmetric WZW models, here we will restrict our
attention only to those particle models which arise from reducing the
two-dimensional WZW models to one dimension. For ease of comparison, the
one-dimensional particle model which arises from dimensional reduction of
the two-dimensional $(p,q)$ supersymmetric WZW model
($p,q = 0,1,2,\dots)$ will be called the ``$(p,q)$ superparticle''.

For simplicity, in this paper we will restrict ourselves to considering
a simple, compact, simply-connected Lie group $G$, with
generators $t^a$, $a = 1,2,\dots,{\rm dim}G$, satisfying
$[t^a,t^b]={f^{ab}}_c$, with ${f^{ab}}_c$ the structure constants of $G$.
We also have $\Tr [t_a t_b] = \delta_{ab}$.
The group manifold has
coordinates $X^i$, $i = 1,2,\dots,{\rm dim}G$. We denote by $g_{ij}$
the Cartan-Killing metric on $G$.
For group elements $h$, we define the one-forms $l=h^{-1}dh$, satisfying
$l^a_{[i,j]} = \half{f^a}_{bc}
l^b_i l^c_j$, $[l^a,l^b]^i = {f^{ab}}_{c}l^{ci}$, and $g_{ij}l^i_al^j_b
=\delta_{ab}$.
The torsion on the group manifold
is given by $H_{ijk} = f_{abc}l^a_i l^b_j l^c_k$, and is given locally
in terms of an antisymmetric tensor potential $b_{ij}$ by $H_{ijk} =
3\partial_{[i}b_{jk]}$.

\subsection{The $(1,0)$ Superparticle}
We begin by considering the `$(1,0)$' superparticle moving
upon a group manifold.
The superspace for this particle has coordinates
$(x,\theta)$, and we define the covariant derivative
\begin{equation}
 D \equiv  \partial_\theta + \theta \partial_t
   = \frac{\partial}{\partial \theta} + \theta \frac{\partial}
   {\partial t}.
\end{equation}
Our superfields $\Phi^i(t,\theta)$ will be maps from the
superspace into the group $G$, with a component expansion
\begin{equation}
        \Phi^i(t,\theta) = X^i(t) + \theta\psi^i(t).
\end{equation}
The action
for the $(1,0)$ superparticle on a group manifold is then
\begin{equation}\label{wobbly}
 S = \int \! dt \, d\theta \, \left(g_{ij} (\Phi) - b_{ij} (\Phi)\right)
         D\Phi^i\partial_t \Phi^j.
\end{equation}

The action (\ref{wobbly}) can be conveniently written in terms of
maps $G = G(t,\theta)$
from the superspace into the group manifold. This form of the action is
\begin{equation}
S = \Tr\int \! dt \, d\theta \, \left((DGG^{-1})
 (\dot{G}G^{-1}) -
  \int^1_0 ds \, \bar{G}^{-1} \frac{\partial \bar{G}}{\partial s}
 \left[\bar{G}^{-1}D\bar{G},\bar{G}^{-1}\dot{\bar{G}}\right]\right) ,
\end{equation}
where $\bar G$ is an extension of $G$ to a map from the product
of the interval $[0,1]$ with the superspace, to the group $G$.
We impose the boundary conditions $\bar{G}\vert_{s=1}=G$,
$\bar{G}\vert_{s=0}=e$, where $e$ is the identity element of the group $G$.
We introduce a matrix
representation for $G$, and all multiplications are taken to be matrix
multiplications.

The equation of motion obtained from the action (5) is
\begin{equation}
 D(G^{-1}\dot{G}) = 0.
\end{equation}
The solution to this equation is
\begin{equation}
 G = U(\theta) e^{t\hat a} V,
\end{equation}
where $U$ and $\hat a$ are time independent, with $U$ a superfield group
element
and $\hat a$ an element of the Lie algebra ${\rm Lie}G$. The parameterisation
(7) of the solutions of the equation of motion is invariant
under the replacements $U\rightarrow Uh$, $\hat a\rightarrow h^{-1}\hat ah$,
$V\rightarrow hV$, where $h$ is any element of the group $G$. This symmetry
can be fixed by choosing (for example) $V = e$.

We now turn to consider the phase space of this system.
There are two ways in which to define the phase space of a classical
system. The first is the Hamiltonian definition as the space of
positions and their conjugate momenta. A second definition of the phase space,
here denoted $P_C$,
is as the space of solutions of the Euler-Lagrange equations of
motion. We will consider the space $P_C$, and describe it as the `covariant
phase space'.
{}From the explicit solution (7) above, taking into account the
gauge-fixing, we see that this space is just $G\times {\rm Lie}G\times\fff$,
where $\fff$ is the space of anticommuting variables $\psi$.
In order to deduce the Poisson brackets of the theory,
we seek on the phase space $P_C$ a differential two-form
which is closed, time-independent, and supersymmetric. The unique such
form is
\begin{equation}
 \omega = \Tr\int \! d\theta \,\left( \delta G \delta(G^{-1}
 D{G}G^{-1})\right)  - \frac{1}{2}\Tr\left(
\delta(G^{-1}D{G})\delta(G^{-1}D{G})
   \right) |_
 {\theta = 0}.
\end{equation}
It will simplify our discussion to use the variable $a\equiv \hat a + \psi^2$
instead of $\hat a$.
Writing $\omega$ in terms of the superfield component fields
 $\{u,a,\psi\}$ then gives
\begin{eqnarray}\label{sympl}
 \omega &=& \omega(u,a) + \frac{1}{2}\Tr\left(\delta \psi \delta \psi\right) \\
    \omega(u,a)  &\equiv& \Tr\left((u^{-1} \delta u) \delta a +
     (u^{-1} \delta u)^2 a\right).
\end{eqnarray}
One can check that this form $\omega$ is invariant under the supersymmetry
transformations $\Delta u = \epsilon u\psi, \Delta\psi = \epsilon(\hat a
-\psi^2),
\Delta\hat a = 0$,
with $\epsilon$ a constant anticommuting parameter.
The symplectic form (\ref{sympl})
factorises into fermionic and bosonic sectors, and we
can invert each sector separately in order to find the Poisson brackets. (That
this factorisation must occur follows from the fact that a redefinition
of the component fields $\psi^i$ in the action (4) decouples the fermions.
We will be presenting here new canonical formulations of the superspace WZW
models, which provide a basis for the calculation of the Poisson brackets of
arbitrary superfield group elements. Hence we will work with the superfield
group
elements $G$, rather than the component field coordinates $X$ and $\psi$, and
will
express $G$ in terms of our new parameterisations of the solutions of
the supersymmetric WZW models.)

After converting to group manifold
coordinates, inverting the bosonic form follows ref. \cite{papasptwo},
and using this
we obtain the fundamental Poisson brackets for the $(1,0)$ superparticle on
a group manifold
\begin{equation}
 \{X^i,X^j\} = 0, \,\,\, \{X^i, a_b\} = R^i_b, \,\,\, \{a_a,
                a_b\} = f_{ab}^c a_c,
\end{equation}
and
\begin{equation}
 \{\psi^i,\psi^j\} = \delta^{ij},
\end{equation}
If we use group elements $u$ rather than coordinates $X$, then the
brackets (10) become
\begin{equation}\label{twelve}
 \{u,u\} = 0, \,\,\, \{u, a_b\} = -ut_b, \,\,\, \{a_a,
                a_b\} = {f_{ab}}^c a_c.
\end{equation}
In order to simplify calculations as well as the
appearance of equations, we will henceforth adopt a condensed notation.
We define elements of the tensor product space $G\otimes G$ as
$u_1 = u\otimes e$ and $u_2 = e\otimes u$. We similarly
 define $t^a_1 = t^a\otimes 1, t^a_2 = 1 \otimes t^a$
and define the Casimir tensor $C_{12} = t^a_1t_{2a} = t^a\otimes t_a$.
In this notation,
equation (\ref{twelve}) may be written
\begin{equation}\label{poissons}
 \{u_1,u_2\} = 0, \,\,\, \{u_1, a_2\} = -u_1C_{12}, \,\,\, \{a_1,
                a_2\} = -[C_{12},a_2].
\end{equation}
(Note that $[C_{12},a_1 + a_2] = 0$.)

Now we will use these fundamental Poisson brackets to
show that the supercurrents of this model satisfy the particle version of
the $(1,0)$ super
Ka\v c-Moody algebra. The currents of the model are $J_L=G^{-1}\dot{G}$ and
$J_R = DG\,G^{-1}$, satisfying the conservation laws $DJ_L=0$ and
$\partial_tJ_R=0$. In the gauge $V = e$ these currents take the form
\begin{eqnarray}
 J_L &=& \hat a = a - \psi^2, \nonumber \\
 J_R &=&  u (\psi + \theta a)u^{-1}.
\end{eqnarray}
A straightforward calculation, using the brackets (\ref{poissons}),
then yields the current superalgebra
\begin{eqnarray}
    \{J_{L1},J_{L2}\} &=& -\left[C_{12},J_{L2}\right], \nonumber\\
    \{J_{L1},J_{R2}\} &=& 0, \nonumber\\
    \{J_{R1},J_{R2}\} &=& C_{12} + \left[C_{12}, \theta_1J_{R2} +
\theta_2J_{R1}
         +\theta_1\theta_2\partial_{\theta_2}J_{R2}\right], \nonumber \\
\end{eqnarray}
which is the particle version of the classical $(1,0)$ super Ka\v c-Moody
algebra
({\it i.e.}, the zero-mode subalgebra).

\subsection{The $(1,1)$ Superparticle}
The $(1,1)$ particle  has superspace coordinates $(x,\theta^+,\theta^-)$,
and we define the covariant derivatives $D_\pm \equiv \partial_{\theta^\pm} \pm
\theta^\pm \partial_t$. The superfields $\Phi^i(t,\theta^+,\theta^-)$
are maps from the superspace into the group $G$.
The action, written in terms of the group elements $G(t,\theta^+,\theta^-)$
corresponding to the coordinates $\Phi^i$, is
\begin{equation}
S = \Tr\!\!\int \!\! dt \, d^2\theta\left( (D_+GG^{-1})
 (D_-G G^{-1}) -
 \int^1_0 \!\!ds \,( \bar G^{-1} \frac{\partial \bar G}{\partial s}
 [\bar G^{-1} D_+\bar G,\bar G^{-1} D_-\bar G] \right).
\end{equation}
($\bar G$ is defined in the same way as in the $(1,0)$ case above.)
The equation of motion is
\begin{equation}
 D_+[G^{-1}D_-G] = 0,
\end{equation}
for which the general solution is
\begin{equation}
 G = U(\theta^+)e^{t\hat a}V(\theta^-),
\end{equation}
with $U$ and $V$ group elements and $\hat a$ in the Lie algebra of $G$.
We will
expand $U(\theta^+) = u(1 + \theta^+ \psi_+)$ and $V=(1+\theta^-\psi_-)v$, with
$u,v\in G$ and $\psi_\pm\in{\rm Lie}G$.

The symplectic form of this model is readily derived, and written in terms of
the
parameters $(u,v,\hat a,\psi_\pm)$ of the solutions, this
form becomes
\begin{equation}\label{blah}
      \omega = \omega(u,\hat a+\psi_+^2) - \omega(v^{-1},\hat a-\psi_-^2)
      + \half \Tr(\delta\psi_+\delta\psi_+  - \delta\psi_-\delta\psi_-),
\end{equation}
where
$\omega(u,a)$ is defined in equation (10). The form given in
equation (\ref{blah}) is degenerate along the directions of the gauge symmetry
$u\rightarrow uh$, $\hat a\rightarrow h^{-1}\hat ah$,
$v\rightarrow hv$, $\psi_\pm\rightarrow h^{-1}\psi_+h$,
where $h$ is any element of the group $G$. We may fix this symmetry
and eliminate the degeneracy by choosing $v=e$. (Note, however, that this
gauge fixing breaks explicit supersymmetry.) In this gauge the symplectic form
is
\begin{eqnarray}
  \omega = \Tr\left((u^{-1}\delta u) \delta a +
              (u^{-1}\delta u)^2 a
           + \frac{1}{2} \left(\delta \psi_+ \delta \psi_+ - \delta \psi_-
             \delta \psi_-\right)\right),
\end{eqnarray}
where $a \equiv \hat a +\psi_+^2$.
The bosonic sector Poisson brackets are then
the same as in the $(1, 0)$ case, and the fermionic sector brackets
are
\begin{equation}
 \{\psi_{1+},\psi_{2+}\} = C_{12}, \,\,\,
 \{\psi_{1-},\psi_{2-}\} = -C_{12},
\end{equation}
with other Poisson brackets vanishing.
The covariant phase space
$P_C$ has coordinates $(u,a,\psi_+,\psi_-)$,
and equals $G\times {\rm Lie}G\times\fff\times\fff$.

The supercurrents of this model are
$J_L=G^{-1}D_-{G}$ and
$J_R = D_+G\,G^{-1}$, satisfying the conservation laws $D_+J_L=0$ and
$D_-J_R=0$. In terms of the covariant phase space coordinates,
in the gauge $v=e$, these currents take the form
\begin{eqnarray}
 J_L &=& \psi_- - \theta^-(a - \psi_+^2 + \psi_-^2), \nonumber \\
 J_R &=&  u (\psi_+ + \theta^+a)u^{-1}.
\end{eqnarray}
A straightforward calculation, using the brackets (23) then yields the
current algebra
\begin{eqnarray}
    \{J_{R1},J_{R2}\} &=& C_{12} + \left[C_{12}, \theta_1^+J_{R2}
          + \theta_2^+J_{R1}
      +\theta_1^+\theta_2^+ \partial_{\theta_2^+}J_{R2} \right], \nonumber \\
    \{J_{L1},J_{R2}\} &=& 0, \nonumber \\
    \{J_{L1},J_{L2}\} &=& -C_{12} + \left[
         C_{12}, \theta_1^-J_{L2} + \theta_2^-J_{L1}
         +\theta_1^-\theta_2^-\partial_{\theta_2^-}J_{L2}\right].
\end{eqnarray}
which is the particle version of the classical $(1,1)$ super Ka\v c-Moody
algebra.

\subsection{The $(2,0)$ Superparticle}
This particle model is that obtained from a dimensional reduction of the
$(2,0)$ WZW model. The latter is defined only on manifolds admitting a suitable
complex structure (the classical formulation of the $(2,0)$ WZW model is given
in
reference \cite{hullsp}, and the quantum theory is discussed in
\cite{hullsptwo}.
The reader is referred to these references for more discussion on the
formulation of this model.)
The particle
superspace has coordinates $(t,\theta^+,\bar\theta^+)$, with supercovariant
derivatives $D_+ = \partial/\partial\theta^+ + \bar\theta^+\partial_t$,
$\bar D_+ = \partial/\partial\bar\theta^+ + \theta^+\partial_t$.
The complex structure on the group manifold may be used to separate the group
manifold coordinate superfield components $\phi^M, M = 1,\dots,{\rm dim}G$,
into
two sets $\phi^m, \phi^{\bar m}, (m, \bar m = 1,\dots,\half{\rm dim}G)$. These
coordinate superfields are required to satisfy the chiral constraints
\begin{equation}\label{constraints}
\bar D_+\phi^m = 0 = D_+\phi^{\bar m}.
\end{equation}
The equations of motion of the $(2,0)$ superparticle are
\begin{equation}\label{twozero}
       D_+(G^{-1}\dot G) = 0 = \bar D_+(G^{-1}\dot G),
\end{equation}
where $G(t,\theta^+,\bar\theta^+)$ are maps from the superspace into the group.
The equations (\ref{twozero}) are solved by
\begin{equation}\label{solutions}
    G = U(\theta^+,\bar\theta^+)e^{ta}v,
\end{equation}
where $U$ is a superfield group element, $v$ a group
element, and $a\in{\rm Lie}G$. We will expand $U$ in component fields as
$U = u(1 + \theta^+\psi + \bar\theta^+\bar\psi + \theta\bar\theta b)$.
The conditions (\ref{constraints}) are then solved if we impose
the following consistency conditions
upon these component fields
\begin{eqnarray}\label{relations}
\psi^{\bar c} &=& 0 = \bar\psi^c,\quad u = e, \nonumber \\
  b^c &=& a^c + (\bar\psi\psi)^c,   \nonumber \\
   b^{\bar c} &=& -a^{\bar c} - (\psi\bar\psi)^{\bar c}.
\end{eqnarray}
(It will turn out that products of the Lie algebra-valued fields $\psi,
\bar\psi$
only appear in the form of commutators, when the above expressions for the
$b$ field components are substituted in expressions below.)
Note that in this model there is no redundancy in the parameterisation of the
solutions, due to the constraints (\ref{constraints}) imposing the relations
(\ref{relations}).

To find the Poisson
brackets of this model, it is easiest to note that the $(2,0)$ model is just
a $(1,0)$ model defined upon a special manifold (i.e., one admitting a suitable
complex structure). Hence, on such a manifold the symplectic form, and the
consequent Poisson brackets,
become those of the $(1,0)$ model. In this case this is just
$\omega = -\omega(v^{-1},a) + \half
\Tr(\delta\Psi\delta\Psi)$, where $\Psi = (\psi^a,\bar\psi^{\bar a})$. Using
$2\omega$ rather than $\omega$ for later convenience, we are thus
led to the Poisson brackets
\begin{eqnarray}\label{moremore}
                \{v,v\} &=& 0, \quad \{v,a^c\} = \half t^cv,\quad
          \{v,a^{\bar c}\} = \half t^{\bar c}v,\nonumber \\
    \quad \{a^b,a^{\bar c}\} &=& \half{f^{b\bar c}}_{\bar d}a^{\bar d}
   + \half{f^{b\bar c}}_da^d,\nonumber \\
      \quad \{a^{\bar b},a^{\bar c}\} &=& \half{f^{\bar b\bar c}}_{\bar
d}a^{\bar d},
   \quad \{a^b,a^c\} = \half{f^{bc}}_da^d,\nonumber \\
       \quad \{\psi^b,\bar\psi^{\bar c}\} &=& -\delta^{b\bar c},
\end{eqnarray}
with the other Poisson brackets vanishing.

The currents of this model are
$J_L = G^{-1}\dot G$ and $J_R = D_+G\, G^{-1}, \bar J_R = \bar D_+G\, G^{-1}$.
(The right currents automatically satisfy the constraints discussed in refs.
\cite{hullsp}, \cite{hullsptwo}, which are $D_+J_R = J_R^2$ and its conjugate.)
The left current is simply $v^{-1}av$ in
the covariant phase space coordinates, and its Poisson bracket with itself
just gives a copy of the Lie algebra.
The right currents are given explicitly by
\begin{eqnarray}\label{rightcurrents}
     J_R &=& \psi + \theta^+\psi^2 + \bar\theta^+(a + b + \psi\bar\psi)
                   + \theta^+\bar\theta^+[a+b + \psi\bar\psi,\psi],
         \nonumber \\
    \bar J_R &=& \bar\psi + \bar\theta^+\bar\psi^2 +
                 \theta^+(a - b + \bar\psi\psi)
                   + \theta^+\bar\theta^+[-a+b-\bar\psi\psi,\bar\psi],
\end{eqnarray}
where the expressions for the components of the field $b$ given in equation
(\ref{relations}) are to be inserted in equation (\ref{rightcurrents}). Note
that
the components $J_R^{\bar a}$ and $\bar J_R^a$ vanish due to the constraints
(\ref{constraints}), as expressed by the relations (\ref{relations}).

Since the expression $v^{-1}av$ has vanishing Poisson brackets with the
variables
$a$, $\psi$ and $\bar\psi$, we see immediately that the left current $J_L$
Poisson commutes with the right currents $J_R, \bar J_R$. The calculation of
the Poisson brackets
of the right currents amongst themselves is rather tedious. However, if we
refer to ref. \cite{hullsptwo}, we note that this calculation is precisely the
classical, particle
analogue of the quantum operator product expansions calculated there.
(The classical calculation corresponds to taking only single operator products
of elementary fields in calculating the operator product of composite fields.
The field redefinitions $\psi^c\rightarrow i\sqrt2\psi^c$,
$\bar\psi^c\rightarrow -i\sqrt2\bar\psi^c$, $j\rightarrow 2a$ relate the
fields of ref. \cite{hullsptwo} to those used here, and the constant $k$ of
ref. \cite{hullsptwo} equals $4i$ here.)
Whence we deduce that the Poisson brackets of the right currents given by
equation (\ref{rightcurrents}) yields the classical, particle version of the
non-linear $N=2$ Ka\v c-Moody algebra presented in ref. \cite{hullsptwo}.
This is ($J$ is $J_R$ in these relations)
\begin{eqnarray}\label{horrendous}
   \{J_1^a,J_2^b\} &=& \bar\theta_{12}{f^{ab}}_cJ_2^c -
\theta_{12}\bar\theta_{12}
                {f^{a}}_{ec}{f^{be}}_d J_2^cJ_2^d, \nonumber \\
    \{\bar J_1^{\bar a},\bar J_2^{\bar b}\} &=& \theta_{12}
        {f^{\bar a\bar b}}_{\bar c}\bar J_2^{\bar c} -
\theta_{12}\bar\theta_{12}
                {f^{\bar a}}_{\bar e\bar c}{f^{\bar b\bar e}}_{\bar d}
            \bar J_2^{\bar c}\bar J_2^{\bar d},  \nonumber \\
      \{J_1^a,J_2^{\bar b}\}  &=& -\delta^{a\bar b} + \theta_{12}{f^{a\bar
b}}_c
                J_2^c +  \bar\theta_{12}{f^{a\bar b}}_{\bar c} J_2^{\bar c}
    \nonumber \\  \qquad &+&
         \theta_{12}\bar\theta_{12}\left( {f^{a\bar b}}_c \bar D_+ J_2^c
           -{f^{\bar b}}_{\bar c\bar e} {f^{a\bar c}}_d J_2^dJ_2^{\bar e}
                \right),
\end{eqnarray}
where $\theta_{12} = \theta_1 - \theta_2$, $\bar\theta_{12} =
\bar\theta_1 - \bar\theta_2$, $J_1 = J(\theta_1,\bar\theta_1)$ and
$J_2 = J(\theta_2,\bar\theta_2)$.
It is interesting that even this classical $(2,0)$ particle algebra is not a
Lie
algebra. That this must occur can be guessed immediately by realising that
the algebra of the right currents must be consistent with the non-linear
constraints, mentioned above, which they must satisfy.


\section{The Supersymmetric WZW Models}

We now turn our attention to the two-dimensional supersymmetric WZW models. We
will work in Minkowski space,
on a cylinder with coordinates $(x,t)$, with $0\leq x\leq 1$,
$x^\pm = \half(x \pm t),$ and $\partial_\pm = \partial_x \pm \partial_t$.

\subsection{The $(1,0)$ WZW Model}
The $(1,0)$ superparticle discussed above is the one-dimensional
reduction of the two-dimensional $(1,0)$ supersymmetric WZW
model. The two-dimensional $(1,0)$ superspace has co-ordinates
$z = (\sigma^{\mu}, \theta^+) = (t,x,\theta^+)$, with $\theta^+$
a real anti-commuting co-ordinate. The supercovariant derivative is
$D_+ = \frac{\partial}{\partial \theta^+} + \theta^+ \partial_+$.
With $\Phi^i(x,t,\theta^+)$ coordinates on the group manifold, the
action for the $(1,0)$ supersymmetric WZW model is
\begin{equation}
S =  \int d^2\sigma \, d\theta^+ \, (g_{ij} - b_{ij})
     D_+ \Phi^i \partial_- \Phi^j.
\end{equation}
Using a group manifold superfield
$G = G(x,t,\theta^+)$ instead of the coordinates $\phi^i$,
the action can be written in the form
\begin{equation}
 S = \Tr\!\int\! d^2\sigma d\theta^+ \!\left( (G^{-1}\partial_-G)
     (G^{-1}D_+G) - \!
    \int_0^1\! ds \, [\bar G^{-1}\partial_-\bar G, \bar G^{-1}\partial_s\bar G]
     \bar G^{-1}D_+\bar G\right),
\end{equation}
where $\bar G(z,s)$ is an extension of the map $G$
to a map from the product of the
superspace with the unit interval, into the group $G$, with boundary
conditions $\bar G(z,0) =  e$ and $\bar G(z,1) = G(z)$.
The equation of motion following from the action (25) is
\begin{equation}\label{feqns}
 D_+(G^{-1}\partial_-G) = 0.
\end{equation}
The left and right currents for this model are given by
$J_L(x^-) =  -G^{-1}\partial_-G$, $J_R(x^+) = D_+G\,G^{-1}$.
The closed, supersymmetric,
time-independent symplectic form for the $(1,0)$ model is found to be
\begin{equation}\label{sform}
 \Omega = \Tr\int_0^1\!dx\left( \int\!d\theta^+ \,\delta G
 \delta(G^{-1}D_+GG^{-1})
  - \frac{1}{2}
       \delta(G^{-1}D_+{G}) \delta(G^{-1}D_+{G}\vert_{\theta^+ = 0})\right).
\end{equation}
The solutions to the field equations
(\ref{feqns}) may be written
\begin{eqnarray}\label{param}
 G(t,x;\theta^+) &=& U(x^+;\theta^+){\em W}(\hat A;x^+,x^-)V(x^-),
 \nonumber \\
 W(\hat A;x^+,x^-) &=& P \, {\rm exp}\! \int_{x^-}^{x^+} \hat A(s) ds,
\end{eqnarray}
with $\hat A$ a $(Lie G)^*$-valued periodic one-form on the real
line, `$P$' denoting path ordering. The superfields $U$ and $V$
are periodic in $x$, and hence so is $G$.
The parameterisation (\ref{param}) of the solutions is invariant under the
transformations
\begin{eqnarray}\label{transf}
 &U(x,\theta^+)& \longrightarrow U(x,\theta^+)h(x), \,\,\,
   V(x,\theta^+) \longrightarrow h^{-1}(x)V(x,\theta^+),
 \nonumber \\
 &\hat A(x)&\longrightarrow - h^{-1}(x)\partial_x h(x) + h^{-1}(x)\hat
A(x)h(x),
\end{eqnarray}
where $h$ is an element of the loop group of $G$, $LG$.
The symplectic form (\ref{sform}) is degenerate along the directions
of the action of the transformations (\ref{transf}).
This symmetry may be fixed by setting $V = e$.
We will expand $U = u(1+\theta^+\psi_+)$.
It is also convenient to use the variable
$A = \hat A - \psi^2$ instead of $\hat A$. Substituting the gauge-fixed
solution into the symplectic form then gives
\begin{equation}
  \Omega =  \Tr\int_0^1\!dx\left(
      \frac{1}{2}(u^{-1}\delta u)\partial_x (u^{-1}\delta u)
            + \half\delta \psi_+ \delta \psi_+  +
               (u^{-1}\delta u)^2 A  +  (u^{-1}\delta u)\delta A\right).
\end{equation}
Following the approach of \cite{papasptwo}, it is straightforward
to invert this form to obtain the Poisson brackets
\begin{eqnarray}\label{pbrsmore}
 \{u_1,u_2\} &=& 0, \,\,\, \{u_1,A_2\} = -u_1C_{12}\delta_{12},
     \,\,\, \nonumber \\
 \{A_1, A_2\} &=& (C_{12}\partial_1 - [C_{12},A_2])\delta_{12}, \nonumber \\
 \{\psi_1, \psi_2 \} &=& C_{12}\delta_{12}, \,\,\,
 \{\psi_1, A_2 \} = 0,
\end{eqnarray}
where $u_1 = u(x_1)\otimes e, A_2 = 1\otimes A(x_2), \partial_1 =
{\partial\over\partial x_1}$, etc., with
$C_{12} = t^a\otimes t_a$ again and $\delta_{12}=\delta (x_1,x_2)$, the delta
function on $S^1$.
In the gauge $V = e$, the WZW currents become
\begin{eqnarray}
 J_L &=& A - \psi^2, \nonumber \\
 J_R &=& u\left(\psi + \theta^+(A + u^{-1}\partial_+ u)\right)u^{-1}.
\end{eqnarray}
Using the fundamental brackets (\ref{pbrsmore}), we deduce the current algebra
\begin{eqnarray}
 \{J_{L1}, J_{L2}\} &=& (C_{12}\partial_1 - [C_{12},J_{L2}])\delta_{12},
\nonumber \\
 \label{eq:br2}
 \{J_{L1}, J_{R2}\} &=& 0, \nonumber \\
 \{J_{R1}, J_{R2}\} &=& (1-\theta^+_1\theta^+_2) C_{12} \,\partial_1\delta_{12}
            + [C_{12}, \theta^+_1J_{R2} + \theta^+_2J_{R1}
         +\theta^+_1\theta^+_2\partial_{\theta^+_2}J_{R1}]\delta_{12},
              \nonumber \\
\end{eqnarray}
which is the classical $(1,0)$ super Ka\v c -Moody algebra.

\subsection{The $(1,1)$ WZW Model}
The analysis of this model follows along the lines of the models considered
above, and the reader may derive it using the above methods.
The equation of motion for the $(1,1)$ model is
\begin{equation}\label{eqmo}
         D_+(G^{-1}D_-G) = 0,
\end{equation}
where $G(x,t,\theta^+,\theta^-)$ is a map from the $(1,1)$ superspace into
the group $G$, and the supercovariant derivatives are given by
$D_\pm = \partial/\partial\theta^\pm + \theta^\pm\partial_\pm$.
The solutions of the equation of motion (\ref{eqmo}) are
\begin{equation}\label{oneone}
       G = u(1+\theta^+\psi_+)\, {\rm P}\,{\rm exp}\,
         (\int_{x^-}^{x^+}\!\!\hat A(s) ds)\, (1+\theta^-
                   \psi_-)v,
\end{equation}
where $u,v \in LG$, $\psi_\pm$ are maps from the circle into ${\bf \Psi}$,
and $\hat A$ is a Lie-algebra valued periodic one-form on the real line. We
will
define $A = \hat A + \psi_+^2$. The gauge symmetry in the parameterisation
(\ref{oneone}) will be fixed by setting $v=e$.

The covariant phase space $P_C$ for the $(1,1)$ supersymmetric WZW model
has coordinates
$(u,A, \theta^+,\theta^-)$, with $u\in LG$, $A$ a $({\rm Lie}G)^*$-valued
periodic one-form on the real line, and $\psi_+,
\psi_-$ anticommuting elements of $L({\rm Lie}G)$.
The fundamental Poisson brackets on $P_C$ are then the brackets for $u$ and $A$
given in equation (\ref{pbrsmore}), together with the further brackets
\begin{equation}
 \{\psi_{+1},\psi_{+2}\} = C_{12}\delta_{12}, \,\,\,
 \{\psi_{-1},\psi_{-2}\} = -C_{12}\delta_{12},
\end{equation}
with all other brackets vanishing. The conserved currents are
$J_L=-G^{-1}D_-G$,
$J_R = D_+G\,G^{-1}$ and
the algebra of these currents may then be shown to
be the classical $(1,1)$ super Ka\v c-Moody algebra
\begin{eqnarray}
    \{J_{L1},J_{L2}\} &=&
-(1-\theta^-_1\theta^-_2)C_{12}\,\partial_1\delta_{12}
                         - \left[C_{12}, \theta_1^-J_{L2} + \theta_2^-J_{L1}
         +\theta_1^-\theta_2^-\partial_{\theta_2^-}J_{L2}\right]\delta_{12}
                ,   \nonumber \\
    \{J_{L1},J_{R2}\} &=& 0, \nonumber \\
     \{J_{R1},J_{R2}\} &=& (1-\theta^+_1\theta^+_2)
C_{12}\,\partial_1\delta_{12} +
      \left[C_{12}, \theta_1^+J_{R2} + \theta_2^+J_{R1}
         +\theta_1^+\theta_2^+\partial_{\theta_2^+}J_{R2}
         \right]\delta_{12}.\nonumber \\
\end{eqnarray}

As we have seen above, the covariant phase spaces of the WZW models considered
all take the form $T^*(LG)\times Y$, where $Y$ is a space of anticommuting
variables. The topology of these covariant phase spaces is concentrated in the
subspace  $T^*(LG)$. Thus the same topological issues which arose
for the bosonic WZW model will occur here. Alternative parameterisations of
the solution spaces of the supersymmetric WZW models which could be considered
in this context are those corresponding  to taking the connection $\hat A$ to
be a constant connection. These parameterisations can be reached from ours by
a partial gauge-fixing, however this fixing encounters the Gribov problem. As
a consequence, these alternative covariant phase spaces have differing
topologies to the ones presented here (they will also have difficulties
associated with choosing a proper gauge-fixing of the redundancy in the
solution parameterisations). The covariant phase spaces presented here have
the advantage that they are diffeomorphic to the spaces of initial data
(this is straightforward to check). These issues are discussed fully in
references
\cite{papasptwo}, \cite{newton}, and this discussion carries over entirely to
the supersymmetric models considered here.

\subsection{Other Supersymmetric WZW Models}
We expect that there are covariant phase space formulations of the WZW models
with more supersymmetries than those studied above. As we have discussed with
regard to the $(2,0)$ superparticle, these models have
superfield group elements which have to satisfy constraints, and because of
this
the analysis involving superfield group elements becomes more involved. A study
of the $(2,0)$ WZW model is possible along these lines, following the analogous
particle discussion given above. WZW models with more
supersymmetries exist - for example, the $N=4$ models exist on the group
manifolds
$SU(2n+1), (n = 1,2,\dots)$ \cite{spindel}, and a similar
covariant phase space analysis should apply in such cases. For these models,
however, due to the constraints
involved a component-field analysis is preferable to a superfield analysis,
although
some simplifications can be realised by working in $N=1$ superspace.


\section{Discussion}
Our solutions of the supersymmetric WZW theories involve an integral
$\int_{x^-}^{x^+}\! ds \hat A(s)$. By choosing some point $x_0\in R$, one
may split this integral into $\int_{x^-}^{x^0}\! ds \hat A(s)
+ \int_{x^0}^{x^+}\! ds \hat A(s)$. Using this, one may write our solutions as
products of chiral fields depending upon only one of the variables
$x^+, x^-$ (note, however, that these chiral fields also depend
upon the arbitrary point $x_0$). One can use our fundamental Poisson brackets
to
study the Poisson brackets of these chiral fields. To do this, one needs to
choose a regularisation carefully, as divergences arise from the delta
functions
in the fundamental Poisson brackets.
This problem has been discussed in a general context
recently in ref. \cite{regularisation}. It would be pleasing if it would be
possible
to define the Poisson brackets of chiral fields in such a way that the
emergence of $r$-matrix relations appears naturally in this approach. One might
expect that zero-mode subtleties would be taken care of by a proper treatment
of the dependence upon the point $x_0$. (These comments also apply to the
bosonic
WZW model.)

The above analysis of the covariant phase spaces of supersymmetric WZW models
makes possible a corresponding discussion of the supersymmetric Toda theories.
This will follow by supersymmetric
Hamiltonian reductions \cite{superduper} of the formulations given here, along
the lines of the bosonic results given in reference \cite{loutoad}.
Furthermore,
it has been found recently that these methods apply to the affine Toda and
conformal affine Toda theories \cite{toappear}, so that the supersymmetric
versions of these theories will also be amenable to this approach.

\section{Acknowledgements}
We would like to thank George Papadopoulos for helpful comments.
G.A. was was supported by an Australian Postgraduate Research Award,
and BS by a QEII Fellowship from the Australian Research Council.

\end{document}